\documentclass[aps,pre,epsfig,floats,twocolumn,superscriptaddress,amssymb,amsmath,floatfix,showpacs]{revtex4-1}
\usepackage{graphicx}
\usepackage{bm}  
\usepackage{amssymb}
\usepackage{amsmath}
\usepackage{color}
\usepackage{amscd}
\usepackage{blindtext}

\begin{document}
\title{Symplectic and antiplectic  waves in an array of beating cilia  attached to a closed body}
\author{Aref Ghorbani}
\affiliation{Department of Physics, Institute for Advanced Studies in Basic Sciences (IASBS), Zanjan 45137-66731, Iran}
\author{Ali Najafi}
\affiliation{Department of Physics, Institute for Advanced Studies in Basic Sciences (IASBS), Zanjan 45137-66731, Iran}
\affiliation{Physics Department, University of Zanjan, Zanjan 313, Iran}
\email{najafi@iasbs.ac.ir}

\date{\today}

\begin{abstract}
By taking into account the  hydrodynamic interactions  in a one dimensional array of model cilia attached to a  no-slip cylinderical surface, 
we investigate their  synchronized motion. 
We show, how does the emergence of metachronal waves depend on the 
initial state of the system and  investigate  the conditions under which, the formation of symplectic and antiplectic  waves are possible. 
\end{abstract}




%
%
%
\maketitle
\section{Introduction}
Cillium, a micron scale flexible hair-like appendix  and their ensembles appear in many biological systems\cite{thecell}.   
Mucociliary transport in respiratory system and swimming of ciliated organisms like Volvox and Paramecium are among the 
most important examples of cilia in biology \cite{blake1971,dummer2016measuring,Cilium2010Force,witzany2016biocommunication,Cilium-Flagella}. 
Using the forces from molecular motor-proteins embeded in its molecular structure, an individual cilium can beat and produce flow field \cite{riedel2007molecular}. 
In most of their natural appearance,  the emergent synchronized motion in the form of metachronal wave developed  in  assemblies of cilia is an essential key in 
their performance. 
This is due to the fact that the flow field corresponding to an individual cilium is negligibly small but a synchronized pattern of ciliary beating is able to either  
produce a net flow of fluid in mucus or generate a swimming mechanism  for the ciliated microorganisms.
The metachronal wave is a kind of synchronized pattern of ciliary beating that results a traveling wave on the envelop of their tips.   
Experimental studies show that the direction of  a metachronal wave can be either parallel (symplectic wave) or antiparallel 
(antiplectic wave) to the direction of  power-stroke in an individual cilia \cite{knight1954relations,witzany2016biocommunication}. 
Physical mechanism behind this wave pattern formation is not completely understood but it is mainly believed that the hydrodynamic interactions between cilia, 
can lead their assembly   to reach a synchronized 
state with propagating  metachronal waves \cite{Cilium1998,lighthill1952squirming,coherentcoupling}. 
There are some experimental  observations in artificial active colloidal systems that support the idea of   hydrodynamic 
mediated synchronization in colloidal systems\cite{Kotar27042010,di2012hydrodynamic,cicuta2,elife}. 
In addition to the hydrodynamic interactions, new studies have suggested  that precise coordination of flagellar motion is provided by contractile 
fibers of the basal membrane \cite{basalcoupling}.
%
In most of recent works a flat geometry for the basal ciliated mebrane has been 
considered \cite{julicher,uchida2011generic,starksync,gomper,pedley,qian2009minimal,golestanian2011hydrodynamic}. 
Motivated from the hydrodynamic effects due to a rough wall \cite{rad}, one can expect to see the effects due to the curvature of a ciliated body in the synchronization 
of its cilia.
In a very recent study, the  synchronization of cilia attached to a sphere has been addressed and it is shown that metachronal waves 
can appear \cite{ciliasphere}. 
In this article we revisit the emergence of metachronal waves on a curved ciliated body 
and consider a ring of  
cilia attached to the peripheral of a cylindrically curved   body. Following the model of Vilfan {\it et al.}, we consider each cilium as a 
small sphere moving along an elliptic trajectory \cite{julicher}. 
To take into account the effect of curvature in the hydrodynamic interaction, we 
use an approximate scheme and assume that the interaction of two adjacent cila can be  calculated using a flat wall that is locally tangent to the surface. 
We will show that as a result of asymmetry in the orbit, both symplectic and antiplectic waves can emerge. 
\begin{figure}
\includegraphics[width=.45\textwidth]{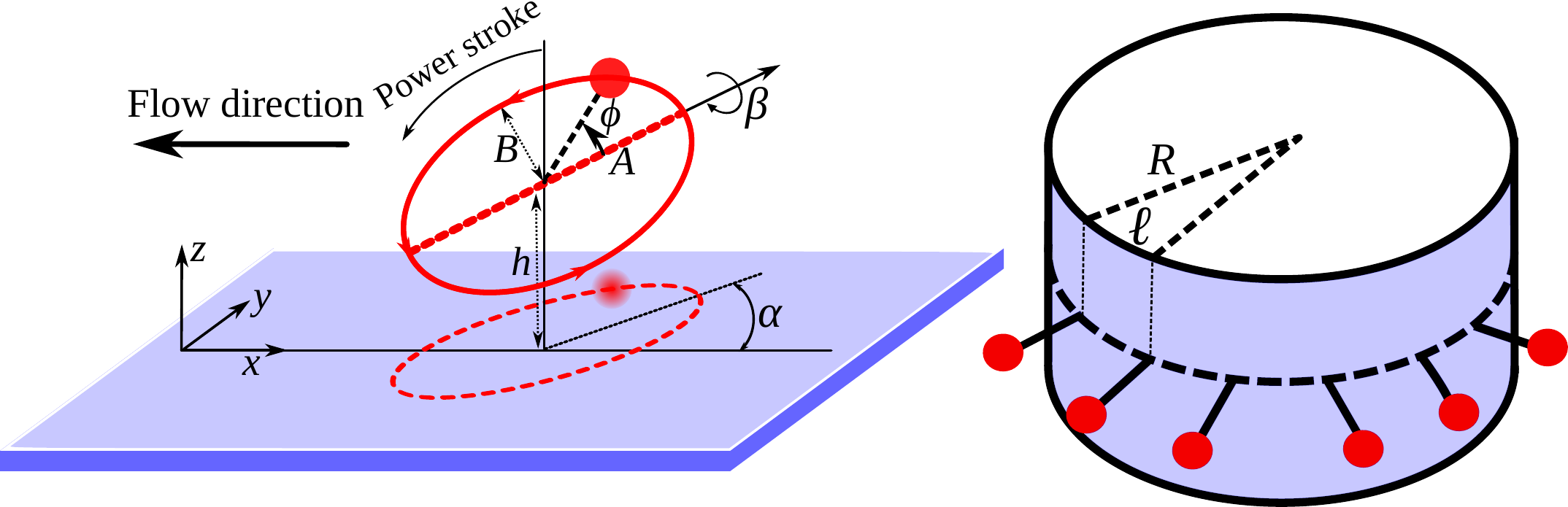}
\caption{Left: geometry of a single cilium, right: an array of cilia attached to the surface of a cylinder. The direction of power stroke and  the average flow are shown in figure.}
\label{fig1}
\end{figure}

\section{Model}
In order to study the motion of an assembly of cilia, we start by defining our simplified  mechanical model for a single cilium. 
To simplify the motion, instead of considering the dynamics of a real cilium which has many degrees of freedom, we can consider the motion of its center of mass. 
Fluid flow produced by a small sphere located at the position of  center of mass resembles the flow pattern due to the cilium. 
Regarding the periodic motion of a cilium, the sphere should move on a closed trajectory.
Fig.~\ref{fig1}-left, mimics the trajectory on which the cilium  center of mass moves. 
Verified by experiments, the ciliary cycle is asymmetric so the friction 
forces are different for the first and second half of the cycle. These half cycles compose the power stroke
and the recovery. This asymmetry that should be reflected in the trajectory, is essential in allowing the cilium to produce a net flow of fluid 
along its stroke direction. 
In determining the dynamics and also the shape of the trajectory,  one should consider the experimental fact that a cilium has a  self sustained dynamics. 
As a result of this self sustained motion the phase variable, angle of the motion along the trajectory, is free. 
This phase freedom is essential in synchronization of two cilia. Considering this phase freedom, two classes of models can be considered. 
In the first class of models it is assumed that the trajectory is on average a circular path. 
This means that the internal forces of molecular motors can be divided into two parts: a constant tangential force  along the  preferred 
trajectory  and an elastic radial restoring force that guarantees an average finite radius  for the trajectory \cite{Lenz,ciliasphere}. 
Such radial elastic force allows the system to behave like a phase-free rotator. 

In the second class of models, instead of fixing a value for the tangential force, its response function, a relation between the  force and 
velocity, has been considered. 
In this article, we will use this kind of modeling to consider the dynamics of cilia \cite{julicher,joanny}. 
Schematic view of the model and its geometrical parameters are shown in fig.~\ref{fig1}(left).   
In a reference frame located on  the wall (Lab. frame), the sphere moves on an elliptic  orbit that is 
characterized by $6$ parameters, $A$, $B$, $h$, $x$, $\alpha$ and $\beta$. 
The lengths of semi-major and semi-minor axises  are denoted by  
$A$ and $B$ and the position vector of the center of ellipse is given by  $(x,0,h)$.  A  rigid wall stands for the body  is placed at  $z=0$. 
The plane of the ellipse and the rigid wall are not parallel, 
the plane of ellipse is rotated  with an angle $\beta$  around its semi-major axis. Projecting the orbit on the plane, the semi-major axis is tilted with 
an angle $\alpha$ with respect to $x-$axis.  
For later use we denote the  eccentricity of the orbit  by  $e=\sqrt{1-(\frac{B}{A})^2}$. 
Instantaneous dynamical state of  the sphere moving on this orbit, is denoted   by an angle  $\phi (t)$. 
In the laboratory frame, the instantaneous  position vector for a cilium that depends on time only through the phase variable $\phi(t)$ can be written as:
\begin{equation}
{\bf r}[\phi(t)]=\begin{bmatrix}
           x \\
           0 \\
           h
         \end{bmatrix}+ {\cal R}_A(\beta){\cal R}_z(\alpha)\begin{bmatrix}
           A\cos\phi \\
           B\sin\phi \\
           0
         \end{bmatrix},
\nonumber
\end{equation}
where ${\cal R}_z(\alpha)$ and ${\cal R}_A(\beta)$ denote the rotation matrices  around $z$ and major axis of the ellipse. In this article we use columnar matrices to show the 
vectors. 

Based on an intuitional argument, we can easily distinguish the 
direction of  average flow produced by a cilium. 
It is essential to note that only  a tilted elliptic  trajectory 
($\beta\neq 0$) is able to produce net flow. For a tilted trajectory,  we can decompose the ciliary cycle into two  sub-trajectories both parallel to the wall, 
one near and the 
other  far from the wall. The cilium has  more or less the same velocity in both parts but the friction coefficient 
is greater in the near wall case. As a result of smaller friction coefficient, the force exerted on fluid is stronger at the part that is far from wall. 
This means that the motion of cilium in a part of its trajectory that is far from the wall, determines the flow direction. 
Thus the flow pattern is in the same direction as the cilium moves  in its motion where it is far from  wall. 
For a typical trajectory shown in fig.~\ref{fig1}(left), the direction of the flow points from right to left. 
In this argument we have neglected the parts of trajectory that are perpendicular to the wall, such parts will have contribution in flow perpendicular to the wall. 
In the case of many coordinated cilia, the perpendicular part of the velocity profile averages to zero. 

In this article we aim to investigate the  curvature of the ciliated body and its role in the dynamic of cilia. In order to attack this problem,  we consider a 
$1$ dimensional  array of ${\cal N}$ cilia attached to a circle 
around the cylinder. The circle is wrapped around the  cylinder and it has the same radius  $R$ as cylinder.   
As shown in fig.~\ref{fig1}(right), two adjacent cilia  are connected with an arc  length $\ell=2\pi R/{\cal N}$. The geometrical 
parameters of each cilia can be expressed with respect to a flat wall that is locally tangent to the cylinder.  
In the case that the length of each cilium, is comparable with this arc-length, we expect to see hydrodynamical effects due to the curvature of body. 
In the next 
section we will summarize all of the  equations that are necessary to describe the  dynamics of a coupled system of cilia. 

\section{Dynamical equations}
Let us consider two cilia, each represented by a moving sphere with radii $a$ and position vectors given by ${\bf r}_i$ ($i=1,~2$) and 
corresponding parameters for their elliptic trajectories. 
Hereafter we consider similar cilia that have same geometrical and dynamical properties. 
At micrometer scale where the dissipative effects dominate over inertial effects, the
governing equations for two interacting  colloidal particles (here two cilia) can be written as  
linear relations between the $i'th$ particle's velocity ${\bf v}_i={\dot {\bf r}}_i$ and the hydrodynamic forces acting on  
particles denoted by ${\bf f}_j$. In terms of their Cartesian components, we have:
\begin{equation}
{v}_{i,\mu}=\sum_{\nu=1}^{3}\sum_{j=1}^{2}G_{\mu\nu}
({\bf r}_i,{\bf r}_j){f}_{j,\nu}
\label{fv}
\end{equation}
where Greek letters denote the cartesian components of the vectors.   
The hydrodynamic kernel $G_{\mu\nu}$ contains information about the geometry  of the system: radii of spheres, their separation and their distances to the wall. 
Assuming that the sphere radius, $a$, is much smaller than all other lengths in system,  we can write an approximate 
form for the component of the hydrodynamic kernel in a semi infinite domain confined by a rigid wall.  For ${\bf r}_i\neq{\bf r}_j$, we have \cite{pozrikidis}:
\begin{equation}
G({\bf r}_i,{\bf r}_i)\simeq \frac{3}{2 \pi \eta} \frac{z_iz_j}{d^3}
\begin{pmatrix}
\cos^2{\psi}&\sin\psi \cos\psi & 0 \\ 
\sin\psi \cos\psi & \sin^2{\psi} & 0 \\ 
0 & 0 & 0
\end{pmatrix},
\end{equation}
where $\eta$ is the fluid viscosity, $z_i={\hat z_i}\cdot{\bf r}_i$, $d=\sqrt{(x_j-x_i)^2+(y_j-y_i)^2}$ and we have assumed that $z_i,z_j\ll d$.
Here $\psi$ is defined as: $\tan \psi=(y_j-y_i)/(x_j-x_i)$. 
For ${\bf r}_i={\bf r}_j$, we have:
\begin{equation}
G({\bf r}_i,{\bf r}_i) \simeq \frac{1}{6\pi\eta a}
\begin{pmatrix}
1-\epsilon& 0 & 0 \\ 
0 & 1-\epsilon & 0 \\ 
0 & 0 & 1-2\epsilon
\end{pmatrix}
\end{equation}
where $\epsilon=(9a/16z)$.
Let us  continue our  discussion about the case of two interacting cilia near a flat wall, then we will discuss how the effects due to the curvature of 
the body can be considered. 
Denoting the inverse of hydrodynamic kernel $G({\bf r}_j-{\bf r}_j)$ by matrices $M_{ij}$, the hydrodynamic equations can be rewritten as:
\begin{eqnarray}
&& {\bf f}_1=M_{11}{\bf v}_1+M_{12}{\bf v}_2\nonumber\\
&& {\bf f}_2=M_{21}{\bf v}_1+M_{22}{\bf v}_2.
\label{force2}
\end{eqnarray}
In addition to the above hydrodynamic equations, we should provide some information about the internal forces inside each cilium that drive 
its beating. 
In addition to constraining  forces that enforce the 
particle to move on  
elliptic orbit, there is also tangential force  that results  the motion  along orbit. 
Denoting the  unit vector tangent to the trajectory of $i$'th cilium by  ${\bf t}_i=\frac{d}{d{\phi}_i}{\bf r}_i$, 
the velocity can be written as: ${\bf v}_i=\dot{\phi}_i{\bf t}_i$. The tangential component of the force reads as: $f_{i}^{t}=\hat{{\bf t}}_{i}^{T}{\bf f}_i$, where 
symbol $T$ denotes the transpose of a columnar matrix and we use 
matrix multiplication rules.  
In general, the tangential force  is related to the velocity of sphere along its trajectory given by $v_{i}^{t}=\hat{{\bf t}}_{i}\cdot{\bf v}_{i}$. 
In linear response regime,  equations like: 
\begin{equation}
{\bf t}_{i}^{T}{\bf f}_i=f_0(1-v_{0}^{-1}|{\bf t}_i|\dot{\phi}_i)
\label{resp2}
\end{equation}
captures the dynamics of $i$'th cilium.
Here $f_0$ is the stall force and it is the amount of external force necessary to 
stop the motion of a beating cilium. A free cilium that is not affected by any external force, moves with velocity  $v_0$. 
In this linear response approximation,  two parameters $f_0$ and  $v_0$ are related to the microscopic details of the cilium. 
For a typical cilium, $f_0\sim 10{\text pN}$, $a\sim 10\mu{\text m}$ and $v_0\sim 100\mu{\text ms}^{-1}$ \cite{joe, Goldestain2012}. 
Using this numerical values, one can have an estimate for 
 the stiffness of the cilium that is defined by:
$\kappa=f_0/6\pi\eta av_0\sim 1$. This stiffness is the only dimensionless parameter that determines the state of motion for a cilium. 
\begin{figure}
\centering
\includegraphics[width=.45\textwidth]{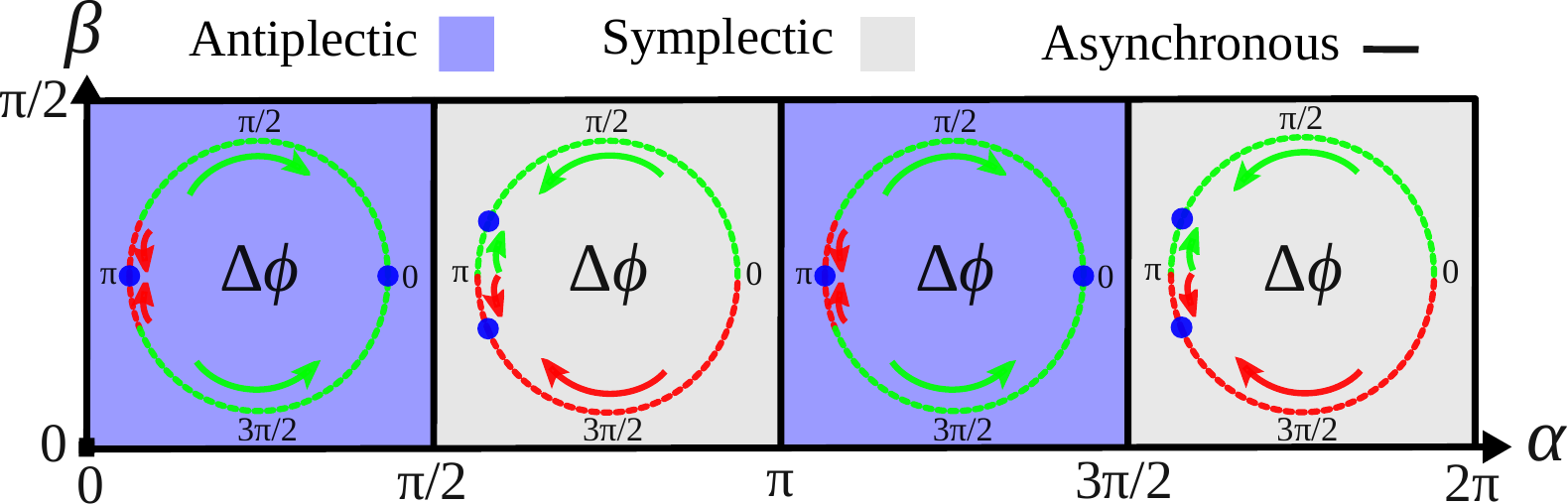}
\caption{Phase diagram for the state of synchronization for two or many cilia. 
Synchronization of two  cilia depends on initial
		phase difference $\Delta \phi(0)$ and $\alpha$. Different values of $\Delta \phi(0)$ are shown by points on a unit  circle. 
		The synchronized states of two cilia are represented by the long time 
		value of their phase differences, $\Delta\phi(\infty)$, and are shown by bolded  dots on the circles. Arrows show the 
		 direction of phase evolution from initial values toward  their final synchronized state, different colors are used to show the evolution to different final states. 
		For many cilia, both symplectic and antiplectic waves can emerge. For any $\beta\neq 0$, antiplectic (symplectic) waves emerge for 
		$0<\alpha<\pi/2$ and $\pi<\alpha<3\pi/2$ ($\pi/2<\alpha<\pi$ and $3\pi/2<\alpha<2\pi$). 
		Numerical parameters are: 
		$a/h=0.2$, $ h/\ell=0.19$, $B/\ell=0.19$, $\beta=\pi/4$ and $e=0.87$. 
		}
\label{fig2}
\end{figure}

Using equations \ref{force2} and \ref{resp2}, we can arrive at the following coupled equations for the phase variables:
\begin{eqnarray}
&&\dot{ \phi}_1\left({\bf t}_{1}^{T}M_{11}{\bf t}_1+(f_0/v_0)|{\bf t}_{1}| \right)+\dot{ \phi}_2{\bf t}_{1}^{T}M_{12}{\bf t}_2=f_0\nonumber\\
&&\dot{ \phi}_1{\bf t}_{2}^{T}M_{21}{\bf t}_1+\dot{ \phi}_2\left({\bf t}_{2}^{T}M_{22}{\bf t}_2+(f_0/v_0)|{\bf t}_{2}| \right)=f_0.\nonumber
\end{eqnarray}
Solving these equations, one can reach to equations that reveals the dynamics of phases for two cilia case. 

Let us now explain how can we take into account the curvature effects. In order to study the curvature, the hydrodynamic kernel 
should be replaced.  
In a confined space that is limited by a curved wall, the above mentioned kernel $G$ and subsequently its inverse given by matrix $M$  is no longer valid. 
We proceed by an approximate scheme to  consider the 
curvature effects. Here we assume that for two interacting cilia, there is an effective plane that can be used for constructing  the  image system. 
This effective wall, is a wall that is locally tangent to the curved body at the mid point of two cilia. 
We can use the results of flat-wall confinement, to obtain  the approximate interaction between two adjacent cilia.   
It is obvious that with this approximation, we do not expect to see any curvature effect in the motion of two cilia. 
This approximation can only include non trivial curvature effects to the motion of many cilia (more than two), attached to the cylinder and 
it allows us to develop a consistent way for applying  closed boundary condition. 
This  approximation is valid  for the case where the radius of curvature is larger than all other length scales in the system namely 
$R\gg \ell$ and $R\gg h$.

To obtain the dynamics of two coupled cilia, one should note that  in a way that we have parametrized the orbits, the kinematics of 
each cilium can be expressed by a single phase denoted by $\phi_i(t)$. This phase variable is shown in fig.~\ref{fig1}.
Solving the dynamical equations and applying the geometrical and dynamical constraints, we will arrive at the following equations \cite{julicher}:
\begin{equation}
\dot{\phi}_i(t)=g_1(\phi_i)\omega_0+g_2(\phi_i,\phi_j).
\end{equation}  
Here $\omega_0=f_0/(6\pi\eta aB(1+\kappa))$ and the interaction between cilia are reflected by function $g_2$ that couples the 
dynamics of phases.  It is not possible to present analytical closed relations for functions $g_1$ and $g_2$ but, in principle we are able to evaluate and study them numerically.  In the next section, we will summarize the results of numerical investigations of  the above equations.

\begin{figure}
\centering
\includegraphics[width=.45\textwidth]{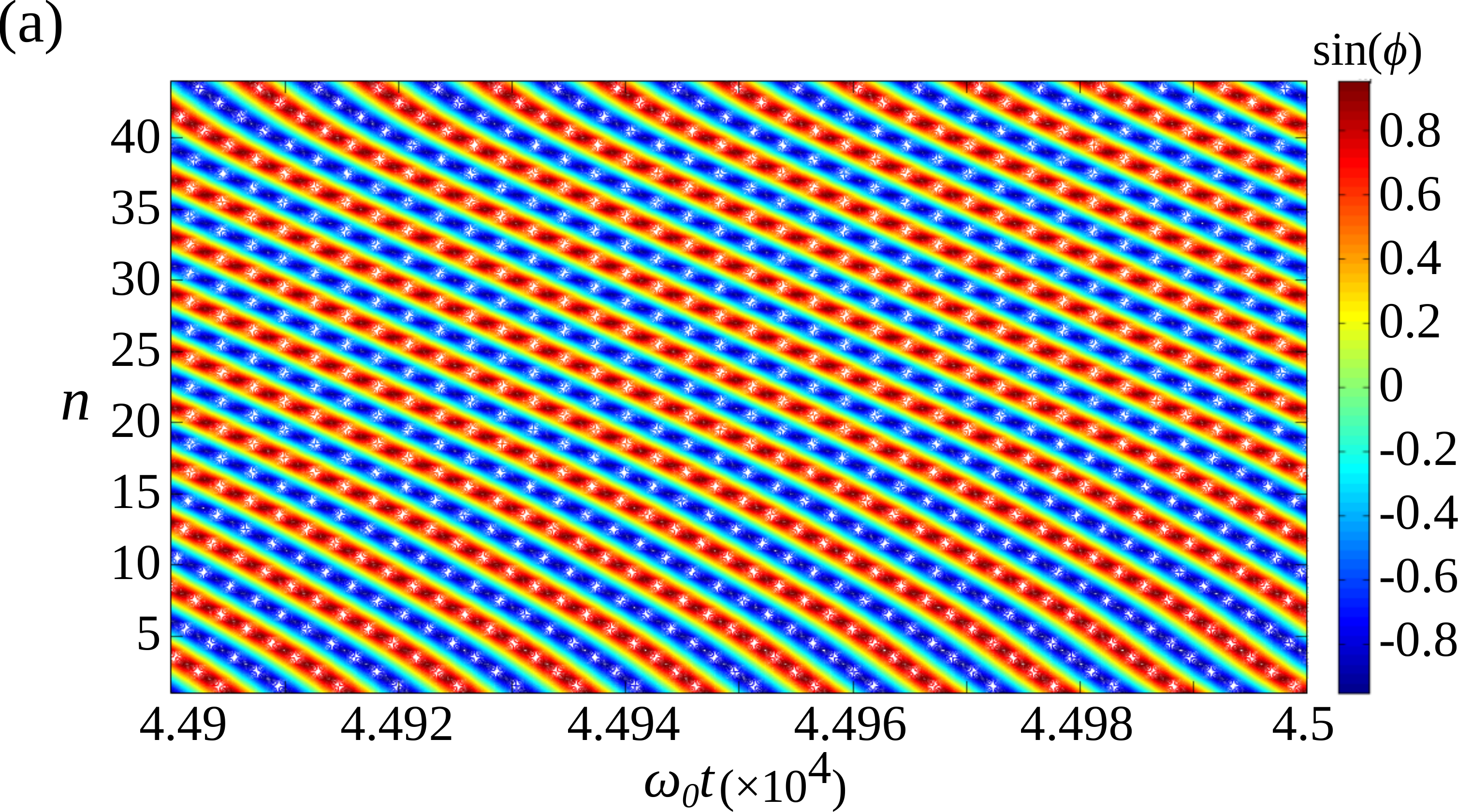}
\includegraphics[width=.45\textwidth]{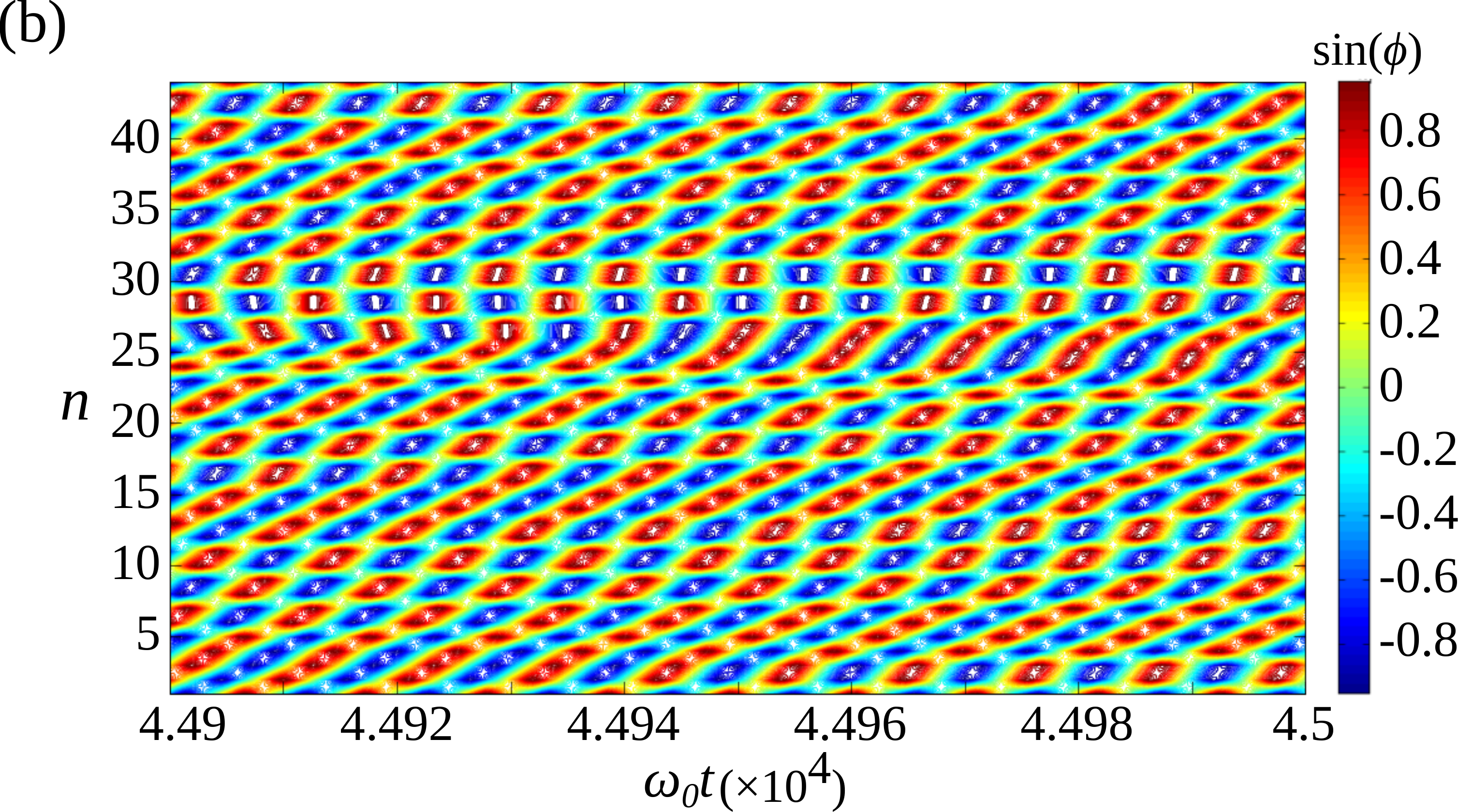}
\caption{State of synchronization in a collection of cilia can be seen in a time-phase portrait. The vertical axes shows   
the cilium  number $n$, the horizontal axis show the time and the value of  $\sin(\phi_n(t))$ are encoded via colors.  
Symplectic wave appears for  $\alpha=2\pi-\frac{\pi}{4}$ (a) and  antiplectic wave appear for $\alpha=\frac{\pi}{4}$ (b). 
The numerical parameters are ${a}/{h}= 0.2$,  $h/\ell=0.35$,  
$B/\ell=0.22$, $e=0.87$  and $\beta={\pi}/{4}$.
}
\label{fig3}
\end{figure}

\section{Results and discussion}
Before studying the wave propagation in ciliated curved body, we  consider the dynamics of two coupled cilia. 
Recalling the geometry of elliptic orbits,  angles $\alpha$ and  
$\beta$, play important role in the dynamics of coupled cilia.  Numerical solutions to the  equations for two interacting cilia,  show that  for $\beta=0$ and $\beta={\pi}/{2}$  (orbits are  parallel and perpendicular to the wall respectively), the long time dynamics of cilia does not  show 
 any correlations in their beating patterns.  In this case, the hydrodynamic mediated interactions between two cilia, 
act in an incoherent way and the state of motion for each cilium is independent from the other.     
 This result is consistent  with previous works of beating cilia near a flat wall \cite{1478-3975-3-4-006,niedermayer2008synchronization}. 
 When elliptic orbits are tilted ($\beta \neq 0,\pi/2$), 
 temporal correlations in  the long time dynamics of interacting  cilia will  appear. 
Our numerical studies show that for tilted orbits,   the cilia reach a phase-locked  synchronized state. 
At this synchronized state, the phase difference $\Delta\phi(t)=\phi_2(t)-\phi_1(t)$ reaches a constant value that we will denote it  by 
$\Delta\phi(\infty)$. The steady state  phase difference, $\Delta \phi(\infty)$,
depends on  initial conditions, the angle of ellipse $\alpha$ and also $e$. 
Fig.~\ref{fig2}, shows the state of synchronization and its dependence to  $\alpha$, $\beta$ and $\Delta\phi(0)$ for $e\neq 0$($=0.87$). 
As one can see from this figure,  for $0<\alpha<\pi/2$ and $\pi<\alpha<3\pi/2$, depending on the initial phase difference between cilia, the final state can 
be either $\Delta\phi(\infty)=0$ or $\Delta\phi(\infty)=\pi$. Behavior for other  values of $\alpha$ and its sensitivity to the 
initial phase difference  can be seen at the figure.  
For $\pi/2 < \alpha < \pi$ and $3\pi/2 < \alpha < 2\pi$, $\Delta\phi(\infty)$ approaches to a constant value, 
$\pi+\delta$, where depending  on the initial phase difference, $\delta$ could be also a positive or negative small angle. 
The angle $\delta$ depends on $\alpha$, as an example for $\alpha=-\pi/4$, it reaches to $\pi/6$.
The synchronization picture shown in fig.~\ref{fig2}, is valid for any $e\neq 0$. 
For a special case of  $e = 0$ where,  the orbits are circular, and  for  $0<\alpha<\pi/2$ and $\pi<\alpha<3\pi/2$, we observed that 
the two  cilia system  reaches a synchronized state with $\Delta\phi(\infty)=0$. When $e=0$ and for  $\pi/2 < \alpha < \pi$ and $3\pi/2 < \alpha < 2\pi$, the 
systems reaches  a state with $\Delta\phi(\infty)=\pi$.

Let us  examine the dynamics of  a one dimensional array of  ${\cal N}=44$ cilia, attached to the circumference 
of a cylinder.   
We take into account the hydrodynamic interactions and considered both cases of elliptic and circular orbits separately ($e=0$ and $e=0.87$).  Interestingly, unlike the case of two cilia, the emergence of synchronized states,   does not show any 
dependence on the value of $e$.   
Fig.~\ref{fig3}, shows two examples for the time evolution of the phase variables for all cilia. The phase values are encoded via colors. 
The  patterns  shown in this figure, demonstrate the long time dynamics of the system.   
The regularity of the long time patterns,  reflects  the  temporal correlations in the motion of cilia. In this case, a traveling metachronal 
wave, shows a synchronized state of the cilia  \cite{Cilium1998,niedermayer2008synchronization,Goldestain2012}.
For a propagating wave (metachronal wave) in the array of cilia, the phase of $n$'th cilium 
$\phi_n(t)$ can 
be written as: $\phi_n(t)=nK-\Omega t$, where the number $K$ plays the role of a wave number associated with metachronal wave. 
In a regular patterns shown in fig.~\ref{fig3}, the regions with constant phases (same colors) make  straight lines. These straight lines are given by an  equation like: $nK=\Omega t+C$, 
where $C$ is a constant. The slope of these parallel lines, measures the wavelength of a metachronal wave and it is  given by $1/K=\partial n/\partial(\Omega t)$. 
For $K>0$, the wave moves in a direction with increasing cilium's number $n$. As shown in fig.~\ref{fig1}, 
the flow direction is always points to the left (from large $n$ to small $n$ cilia). So  we conclude that,  
positive slope corresponds to  antiplectic  and  negative slope shows a symplectic metachronal wave.
Fig.~\ref{fig3}, show two example of metachronal waves for $\alpha=\pi/4$ and $\alpha=2\pi-\pi/4$. 
As one can see, in the first case (fig.~\ref{fig3}(up)), a symplectic wave has appeared, and for the second case (fig.~\ref{fig3}(down)), an antiplectic wave has appeared.
Results of our numerical investigations for the synchronization of an ensemble of cilia are shown in fig.~\ref{fig2}. Results show that, 
the emergence of such  synchronized states (simplectic and antiplectic),  
crucially depends on the value of $\alpha$. 
Independent of the value of $e$, the antiplectic metachronal waves appear for  $0<\alpha<\pi/2$ and $\pi<\alpha<3\pi/2$. On the other hand, 
symplectic metachronal waves appear when $\pi/2 < \alpha < \pi$ and $3\pi/2 < \alpha < 2\pi$. Changing $\alpha$ to 
 $2\pi-\alpha$, the average  direction of the fluid flow does not change but   the propagation direction of the  metachronal wave  will change. 
 Comparing the results for an array of cilia with the results of two cilia, one can see that the  emergence of symplectic or antiplectic synchronization in an array of 
 cilia, is directly related to the  state of synchronization in the case of two cilia.   
 It is interesting that the antipletic wave accompany many defects in their structures. Such defected waves have been seen in a chain of model cilia \cite{cicuta2}.

In conclusion, the state of synchronization is studied for interacting cilia.  For  two cilia, the synchronized state depends on the initial phase difference and the geometrical parameters of the trajectories given  
by  $\alpha$ and $\beta$.  In-phase, anti-phase ($\delta\phi=0,~\pi$) synchronized states have been observed. 
For a $1-D$ array of 
cilia attached to a curved body, we found that,  both symplectic and antiplectic  metachronal waves can appear. The emergence of 
symplectic and antiplectic waves in an array of cilia, is in  direct connection with the state of synchronization in two cilia case. As a result of 
our study, we understand the the  geometrical characteristics are key elements in determining the state of metachronal waves.
The emergence of symplectic or antiplectic metachronism in our model is not an artifact of  imposing closed boundary condition. 
The boundary condition in our system, emerges naturally from the closed structure of the curved body. 

%

\end{document}